# Thickness dependent hydrophobicity of epitaxial graphene

M. Munz[1], C.E. Giusca[1], R.L. Myers-Ward[2], D.K. Gaskill[2], and O. Kazakova[1]

[1]National Physical Laboratory, Hampton Road, Teddington TW11 0LW, UK

[2]U.S. Naval Research Laboratory, Washington, DC 20375 USA

## Abstract

This article addresses the much debated question whether the degree of hydrophobicity of single-layer graphene (1LG) is different from the one of double-layer graphene (2LG). Knowledge of the water affinity of graphene and its spatial variations is critically important as it can affect the graphene properties as well as the performance of graphene devices exposed to humidity. By employing chemical force microscopy (CFM) with a probe rendered hydrophobic by functionalization with octadecyltrichlorosilane (OTS), the adhesion force between the probe and epitaxial graphene on SiC has been measured in deionized water. Owing to the hydrophobic attraction, a larger adhesion force was measured on 2LG domains of graphene surfaces, thus showing that 2LG is more hydrophobic than 1LG. Identification of 1LG and 2LG domains was achieved through Kelvin probe force microscopy and Raman spectral mapping. Approximate values of the adhesion force per OTS molecule have been calculated through contact area analysis. Furthermore, the contrast of friction force images measured in contact mode was reversed to the 1LG/2LG adhesion contrast and its origin was discussed in terms of the likely water depletion over hydrophobic domains as well as deformation in the contact area between AFM tip and 1LG.

*Keywords:* epitaxial graphene; hydrophobic/hydrophilic; chemical force microscopy (CFM); lateral force microscopy (LFM); Kelvin probe force microscopy (KPFM)

Since its first successful isolation via the mechanical exfoliation method[1, 2], graphene, a single atomic sheet of graphite, has attracted an enormous amount of interest, owing to its outstanding electrical, optical and mechanical properties[3]. Major challenges being addressed are the large-scale production of graphene[2, 4] as well as control of its properties under a variety of environmental conditions[5, 6].

The properties of graphene, a two-dimensional (2D) material, are highly susceptible to surface interactions, thus enabling chemical sensing applications with high sensitivity. Adsorption-induced doping[2] can alter the characteristics of a graphene device and make it susceptible to variations of the environmental conditions. In particular, this is the case for operation under ambient conditions, where water molecules or other foreign species[7, 8, 9] can adsorb from air.

The adsorption of water molecules on a graphene surface is closely related to its hydrophobicity. A clean, multi-layer graphene (MLG) surface may be expected to mirror the properties of graphite, which consists of a diverging number ($N \to \infty$) of stacked graphene layers and, a priori, is widely considered non-polar and thus hydrophobic (see Ref. [9] and references therein). However, in the limit of 1LG or 2LG, property variations may occur, e.g., due to doping by the substrate. In particular, wetting transparency was reported for graphene on Cu, Au and Si supporting substrates[10]. Theoretical studies indicated variations in the wetting behavior of 1LG, 2LG and MLG surfaces[11, 12].

Since variations in water wettability are likely to affect graphene properties, ranging from adhesion and friction[13], to electrical charge doping[14] and permeation[15], it is important to develop a thorough understanding of its wetting behavior. Frequently, experimental studies rely on measurements of the water contact angle (WCA) as it provides a measure for the degree of hydrophobicity of graphene surfaces[16, 10, 11, 17, 9]. However, such WCA measurements are typically performed on a macroscopic length scale that averages over a large number of graphene domains. To account for the micron-scale morphology of epitaxial graphene samples, microscopic or mesoscopic techniques are needed. Chemical mapping techniques, such as Raman spectral mapping can be highly useful in identifying different domains of a complex graphene morphology[18, 19, 20], however, complementary techniques are required to measure quantities directly related to the wetting behavior. Chemical force microscopy (CFM), a variant of atomic force microscopy (AFM), uses chemically functionalized probes to measure either force-distance curves or the lateral forces (lateral force microscopy, LFM) occurring between the scanning tip and the surface under investigation[21, 22, 23, 24, 25, 26]. Employing CFM, both the adhesion and friction properties of graphene were studied previously[27, 28, 29].

It should be noted that when measuring ***in air***, specific force interactions can be obscured by the relatively large capillary force resulting from the condensation of water in the narrow gap between AFM and sample surface, including contributions related to the capillary pressure and to the surface tension[30, 31]. At non-zero levels of relative humidity, the capillary force tends to be the major force component[27, 31] and is larger on hydrophilic areas where the water contact angle is smaller[30, 32, 28, 33]. On hydrophilic surfaces, the measured pull-off force largely increases with the relative humidity (RH), although non-monotonic variations may occur in the regime of high RH levels above ~70-80%, depending on the particulars of the AFM tip geometry[30, 31, 33]. In contrast to hydrophilic surfaces, a vanishing RH dependence is typically observed on hydrophobic surfaces[30, 33, 34]. Thus, in the presence of humid air capillary forces play a vital role if the surface is hydrophilic and the measured pull-off force, $F_{po}$, tends to be larger (Fig. 1(a)).

In the case of AFM tip and sample immersed into ***a liquid medium***, the measured adhesion force critically depends on the polarity of the liquid. The adhesion force between hydrophilic surfaces was found to be larger for a hydrophobic liquid[22, 35]. Conversely, the adhesion force measured between two hydrophobic surfaces tends to be larger if they are immersed into a polar liquid, e.g. water. The latter phenomenon is referred to as hydrophobic attraction[36] and has been addressed by several reports[22, 35, 37] considering the effect of the liquid medium polarity on the measured pull-off force, $F_{po}$. Essentially, such studies suggest that in a polar liquid $F_{po}$ scales with the degree of hydrophobicity (Fig. 1(b))[38].

We utilize the hydrophobic attraction for analytical purposes, i.e. to discern between hydrophilic and hydrophobic surface areas through adhesion force measurements (Fig. 1(b)), notwithstanding some controversies[36, 39, 40, 41] concerning the origin of the hydrophobic attraction. We have chosen water as a polar liquid to facilitate comparatively large adhesion forces between hydrophobic AFM tips and domains of epitaxial graphene with different degrees of hydrophobicity.

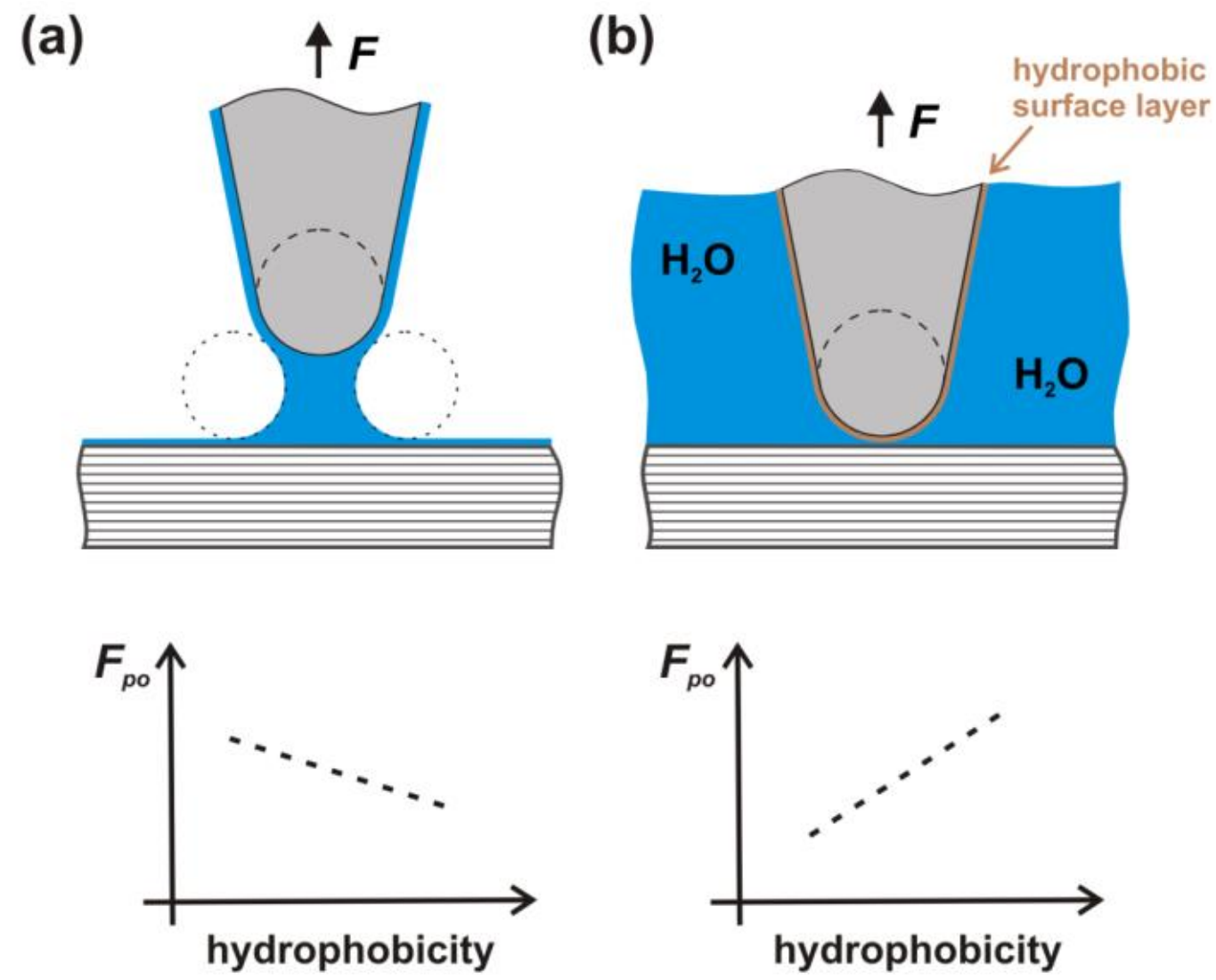

**Figure 1**. Schematic of AFM adhesion force measurements under different environmental conditions and related variations of the pull-off force, $F_{po}$, with the degree of hydrophobicity of the surface under investigation. (a) In humid air, using a plain Si tip. (b) In water, using an AFM tip with a hydrophobic surface functionalization.

In this article, the results of CFM measurements on epitaxial graphene are reported with a focus on the adhesion force variations between 1LG and 2LG. AFM tips with a hydrophobic surface functionalization have been used to map spatial variations in the hydrophobicitiy of epitaxial graphene surfaces synthesized by graphitization of a SiC single crystal. The surface of Si AFM tips was chemically functionalized, using a $CH_3$ terminated monolayer of octadecyltrichlorosilane (OTS, $CH_3(CH_2)_{17}SiCl_3$), and the adhesion mapping was carried out in de-ionized water. Using test samples with a hydrophobic/hydrophilic pattern, the approach relying on hydrophobic attraction has been demonstrated. Its application to the adhesion contrast occurring on 1LG and 2LG domains indicates that 1LG is less hydrophobic than 2LG. Furthermore, this finding has been analyzed in conjunction with subsequent friction force measurements in water and is compared to results of adhesion force measurements conducted under ambient conditions.

**Results and Discussion**

(a) Adhesion and surface potential of ***epitaxial graphene under ambient conditions***

Initially we considered the case of graphene under ambient conditions where capillary condensation occurs. Using Kelvin probe force microscopy (KPFM) mode, we imaged the surface potential of the graphene surface and its spatial variations. The KPFM signal is helpful in quantitatively characterizing the local graphene layer number and identifying 1LG and 2LG domains[42, 16, 43]. The height, adhesion and KPFM images of the graphene surface are given in Fig. 2. The surface potential image (Fig. 2(c)) shows a strong contrast between 1LG and 2LG domains, due to the relative difference in their work functions. This is furthermore highlighted by the histogram (Fig, 2(d)) associated with the area marked with the white square in Fig. 2(c), displaying a bimodal surface potential distribution corresponding to 1LG and 2LG. To determine the surface potential values associated with 1LG and 2LG, peak deconvolution was carried out using Gaussian shape components, and layer thickness was designated in accordance with the domain contrast in the surface potential image.

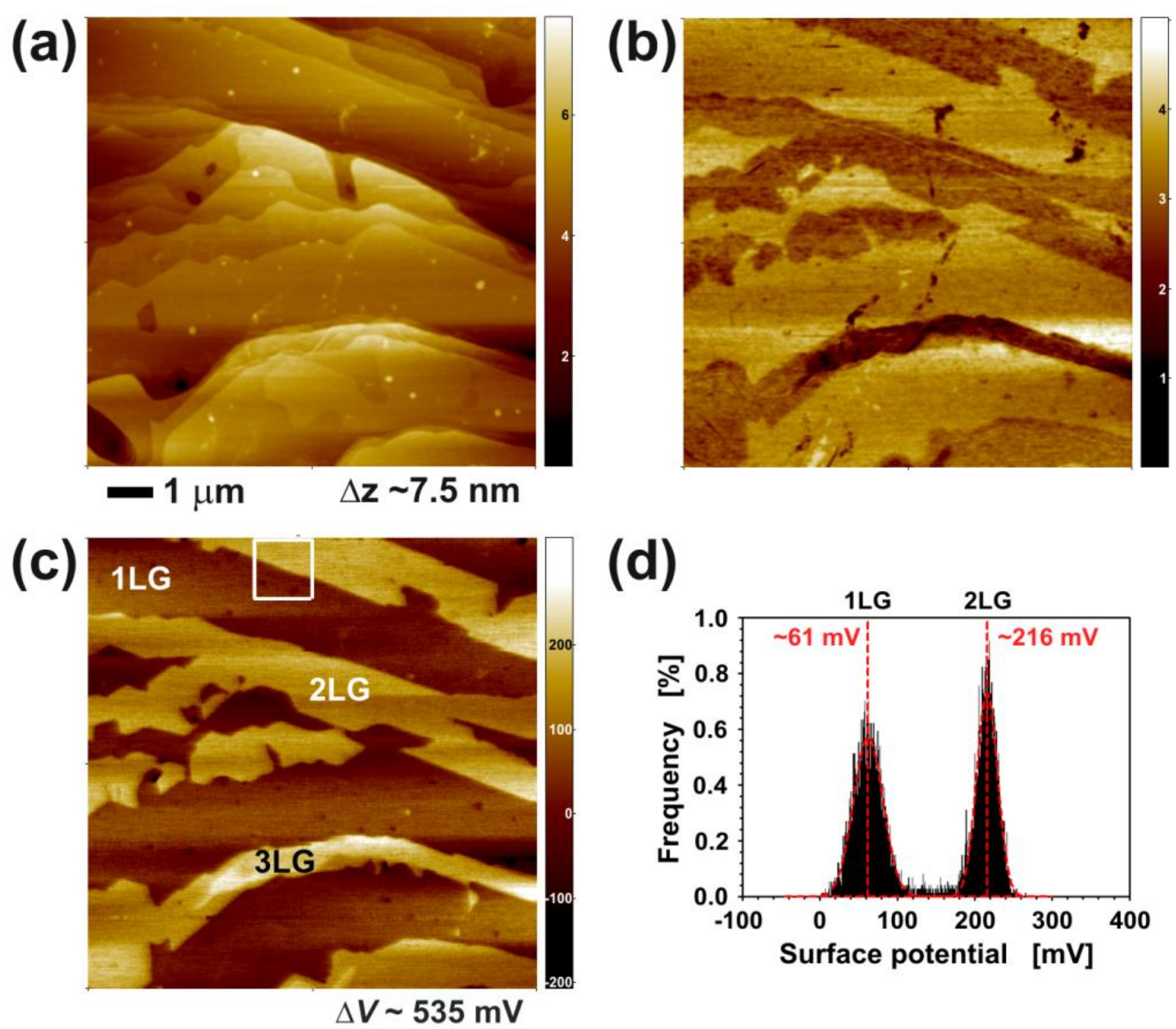

**Figure 2.** Epitaxial graphene surface imaged in air (RH ~40%), using a hydrophilic AFM probe. (a) Topography, (b) adhesion measured in Peak Force mode, (c) surface potential measured in FM-KPFM mode. (d) Histogram associated with the area enclosed by the white square in (c). In (b), the adhesion force range is ~5.1 nN. The scan width in (a-c) is 10 μm.

The surface potential values obtained for 1LG and 2 LG (~61 and 216 mV, respectively) have been used to determine the corresponding work function values, using $eU_{CPD} = \Phi_{tip} - \Phi_{surface}$, where $U_{CPD}$ is the contact potential difference (surface potential) measured by KPFM[43]. The work function of the KPFM tip (4.52 ± 0.02eV) was determined by calibration against a reference gold sample, the work function of which had been measured using ultra-violet photoelectron spectroscopy. A work function value of (4.47 ± 0.02) eV was obtained for 1LG and of (4.32 ± 0.02) eV for 2LG. The value obtained for 1LG is consistent with previous KPFM studies in ambient, reporting a work function value of ~ 4.5 eV for graphene[44, 45, 46]. The assignment of the observed domains to 1LG and 2LG is further corroborated by Raman maps of this sample as detailed in Fig. S1 (*Supplementary Information*), showing a morphology similar to that observed by KPFM.

Furthermore, the adhesion image in Fig. 2(b) shows larger pull-off forces on the domains where a larger work function was measured, i.e. on the 1LG domains. As the measurements were taken in humid air (RH ~ 40%) and the tip was hydrophilic (doped Si tip with no further surface functionalization), capillary condensation of water in the narrow gap between AFM tip and sample surface occurred. Following the rationale depicted in Fig. 1(a), under these conditions a larger pull-off force is observed on areas which are less hydrophobic. Thus, the images in Fig. 2(b) and (c) indicate that 1LG domains are less hydrophobic than 2LG domains.

Additionally, a thin curved domain across the lower half of the images in Fig. 2(b) and (c) can be seen, which is characterized by lower adhesion and higher surface potential values than 1LG and 2LG domains. It can be identified as a 3LG domain that appears even more hydrophobic than the 2LG domains.

(b) Adhesion mapping of test samples ***in water***

To demonstrate the effect of the tip surface chemistry on the tip-sample adhesion force, two test samples were measured in DI water using plain as well as OTS functionalized tips. OTS binds to silica surfaces via the silane group and exhibits a terminal $CH_3$ group at its free end. The structural-mechanical properties of OTS monolayers have been described in several reports[34, 40, 47, 48]. For an OTS monolayer on $SiO_2$, the WCA is ~110° [11, 49]. Furthermore, a typical value for the surface free energy of $CH_3$-terminated monolayers with water is $\gamma_{sl}$~ 52 mJ/m$^2$ [23, 50].

Both test samples consisted of arrays of gold squares on top of a silicon wafer. By making use of the silica surface layer of the silicon wafer, one of the test samples was functionalized with OTS. As a result, the hydrophilic gold squares are surrounded by a hydrophobic area terminated with $CH_3$ groups of the OTS molecules (Fig. 3(b)). In the following, this sample is referred to as test sample TS1. Contrary to the 1$^{st}$ control sample, the 2$^{nd}$ one was treated in an UV/ozone cleaner but no further functionalization was applied. That is, its surface remained hydrophilic throughout, unless adsorption of hydrocarbons from the air occurred. This sample is referred to as test sample TS2. Force mapping results of test sample TS1 are given in Fig. 3 and of sample TS2 in Fig. S2 (*Supplementary Information*). In both cases an OTS functionalized probe was used. As can be seen from the adhesion map shown in Figs. 3(c), a larger adhesion force was measured on the area between the gold squares, i.e. the hydrophobic area terminated with $CH_3$ groups. The histogram related to the force map of Fig. 3(c) is given in Fig. 3(d). It encompasses a distinct peak with its center at ~3.1 nN, and a shoulder on the low force side, which indicates the presence of a second peak with the center around ~1.7 nN. The former is related to the OTS functionalized area, which is the larger fraction of the total area, and the latter is related to the gold squares where the adhesion force was lower. As the maps of Fig. 3 show ~12.5 squares, the relative area covered by the squares is 12.5 μm$^2$ / 100 μm$^2$ ~12.5%.

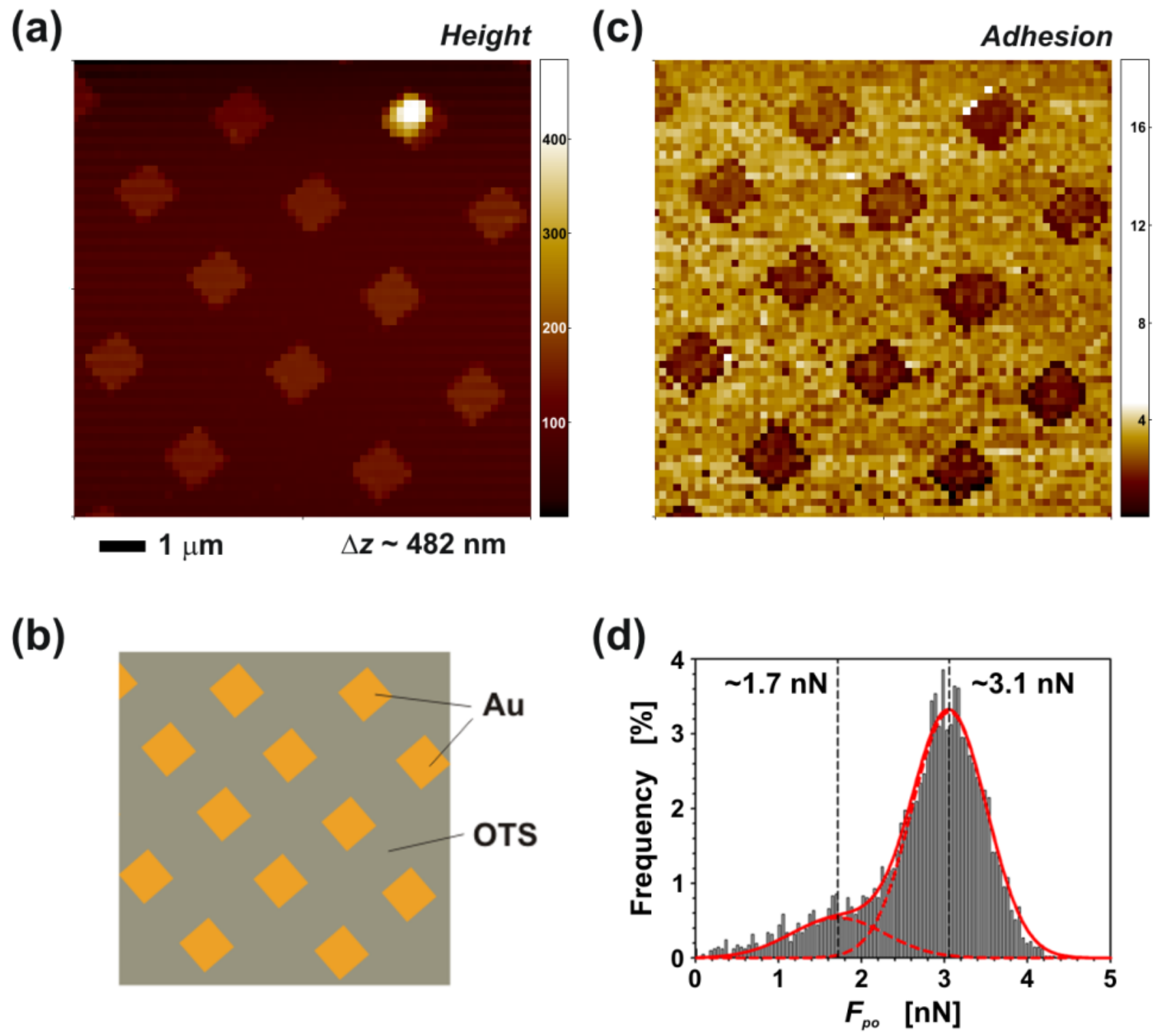

**Figure 3.** Results of force mapping of TS1 (with OTS functionalization), using an OTS functionalized probe. (a) Height map, (b) schematic of the chemical pattern, (c) adhesion map, (d) histogram of the adhesion map with the Gaussian fits to the peaks indicated by red lines. In (c), the adhesion force range is ~18.8 nN. In (a, c), the scan width is 10 μm, the pixel number is 64x64. Cantilever spring constant is ~184 pN/nm, resonance frequency in air is ~20.92 kHz.

Thus, for the present condition of force mapping in DI water, a larger adhesion force was measured for the OTS-OTS contact. This finding is consistent with Fig. 1(b), which illustrates that in water the adhesion force is larger if both surfaces are hydrophobic.

(c) Adhesion mapping of ***epitaxial graphene in water***

Similarly to the test samples, the surface of a graphene sample grown epitaxially on SiC was analyzed in DI water, using OTS functionalized probes. An adhesion map and typical force-distance curves (FDCs) are given in Fig. 4(a). The pull-off peaks occur upon retraction (blue curves), and their height is a measure for the tip-sample adhesion force. Each pixel of the adhesion map gives the absolute value of the peak height (attractive forces are negative). Clearly, the adhesion map shows bright lamellae, where the pull-off force is larger, alternating with darker lamellae. The associated histogram (Fig. 4(b)) shows a bimodal distribution with peaks around ~(4.6 ± 1.4) nN and ~(7.6 ± 1.6) nN, where the uncertainty levels are given by FWHM/2. Following the above rationale (Figs. 1, 3 and 4), areas where larger adhesion forces between the surface and the hydrophobic tip occur are more hydrophobic than areas where the adhesion force is lower. Furthermore, from Raman and KPFM mapping it is known that such lamella-shaped domains are typically 2LG, whereas the areas in between are 1LG (Fig. S1 in the *Supplementary Information* and Fig. 2c). Taken together with the adhesion contrast, this suggests that 2LG is more hydrophobic than 1LG.

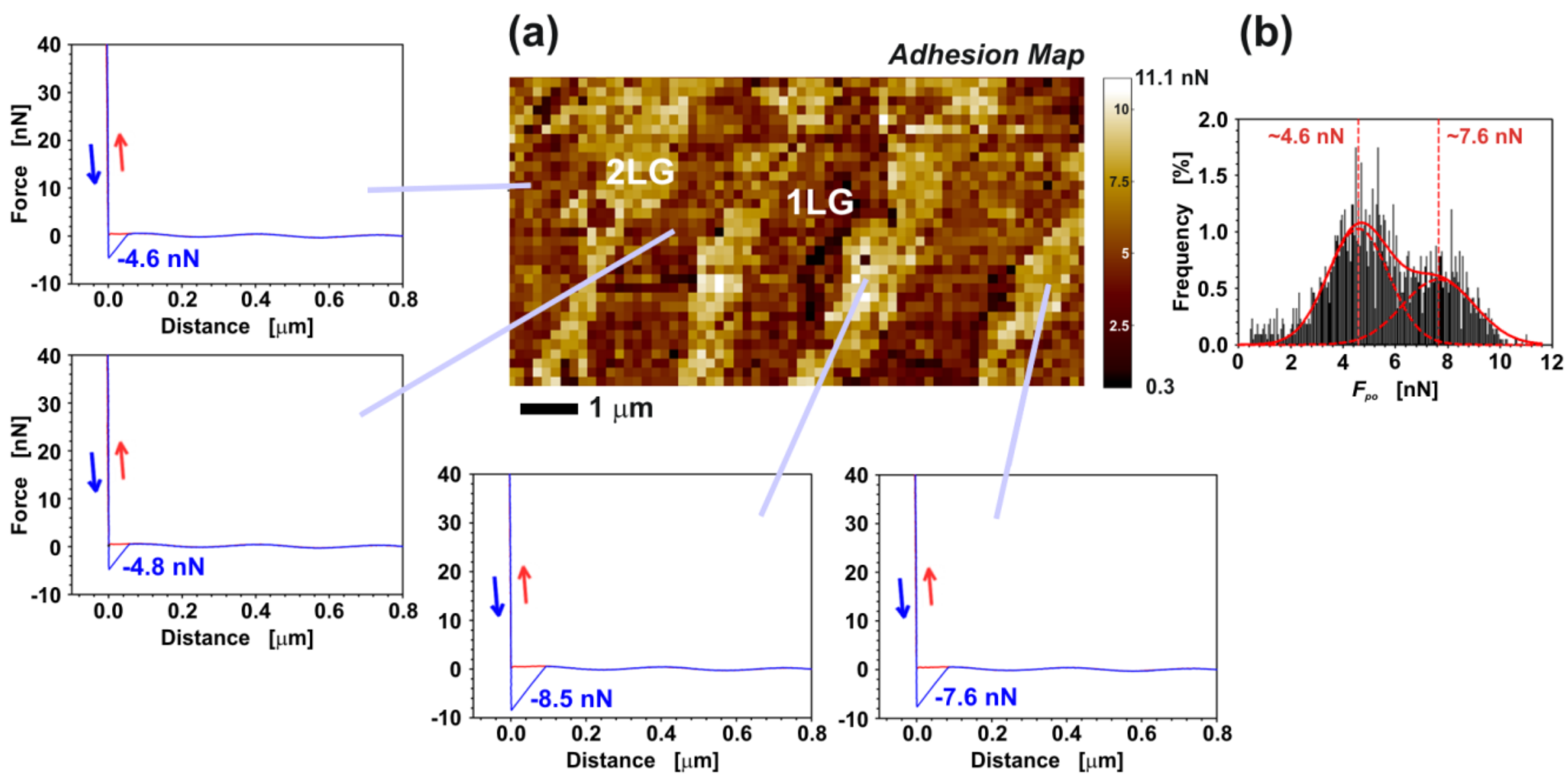

**Figure 4.** (a) Adhesion map of the graphene surface in DI water and typical force-distance curves measured at the indicated locations. The force range of the adhesion map is ~10.8 nN. Force-distance curves measured upon approach and retraction are shown in red and blue, respectively. (b) The histogram associated with the adhesion map shows peaks at ~4.6 and ~7.6 nN. The FWHM values of the Gaussian peaks are ~2.7 and 3.2 nN, respectively.

An adhesion map with a similar contrast but a less regular pattern of domains is given in Fig. 5(a). The associated histogram (Fig. 5(b)) shows a broad distribution which has been deconvoluted by analyzing the histograms (Fig. S4 in the *Supplementary Information*) of both dark and bright adhesion map areas. The analysis yields two Gaussian peaks at ~(1.9 ± 0.4) nN and (2.3 ± 0.3) nN, and the total distribution relating to the entire map can be approximated by a linear combination of these Gaussians.

It should be noted that the measured adhesion force is the total force needed to disrupt the "bonds" between the OTS functionalized AFM tip and the graphene surface. The measured adhesion force is due to the strength of attraction between a group of OTS molecules interacting across the tip-sample contact with the graphene surface, governed by the finite size of the AFM tip. In the following, this interaction is referred to as a "bond", even if no actual covalent bond is formed. A priori, it is unknown whether it corresponds to a single or several "bonds". An estimation of the adhesion force between a single OTS molecule and 1LG or 2LG is given below (Section (d)).

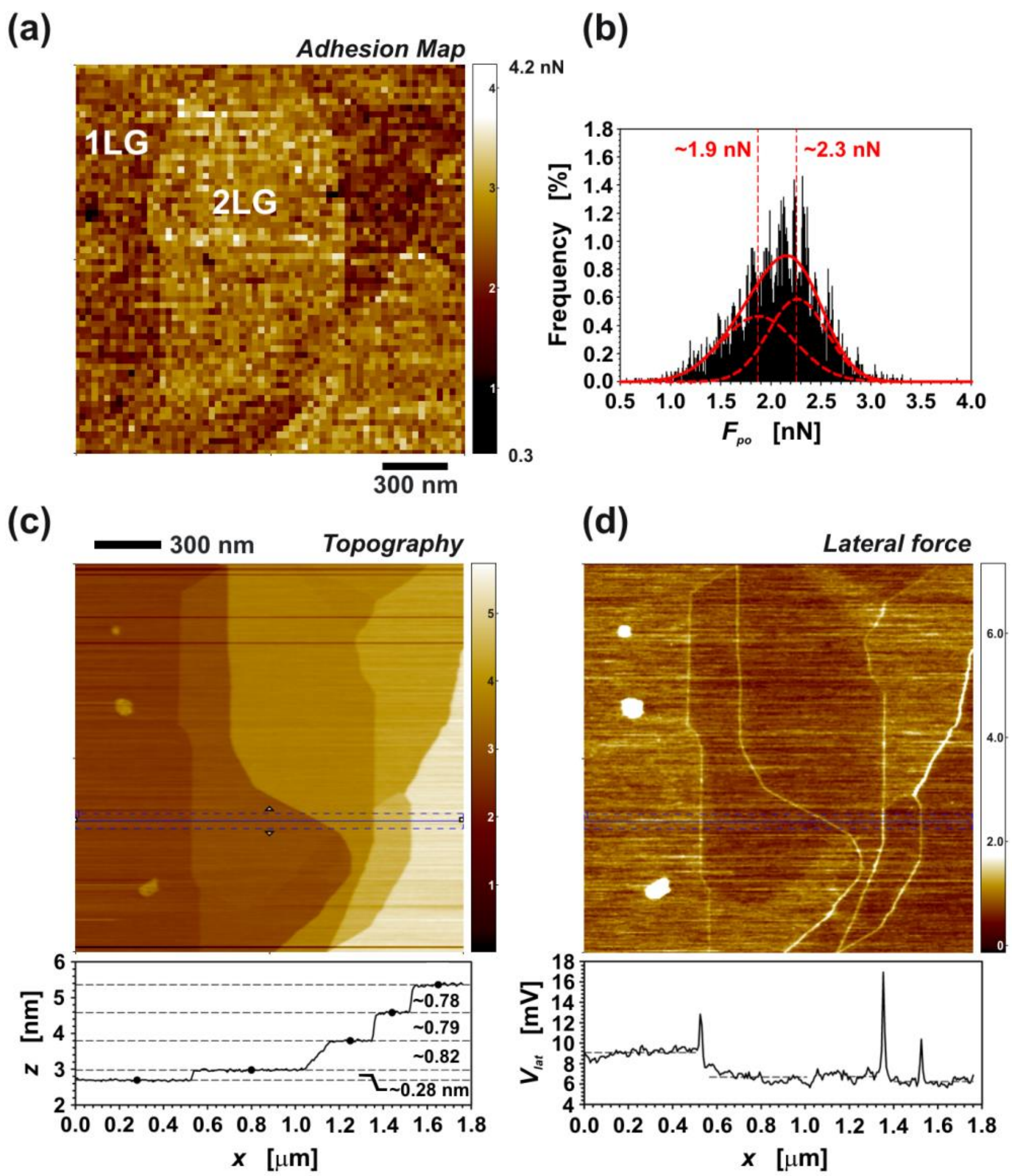

**Figure 5.** Adhesion and friction force measurements on graphene, using an OTS functionalized hydrophobic tip in water. (a) Adhesion map, 64x64 pixels, force range ~3.9 nN. (b) The histogram associated with the adhesion map is composed of peaks around ~1.9 and ~2.3 nN (further analysis of the data is shown in Fig. S5 in the *Supplementary Information*). (c) Topography image measured in contact mode and a cross-sectional profile (averaged over 11

consecutive lines within the horizontal rectangle marked with a dashed blue line). (d) Lateral force image and a cross-sectional profile. In (a), (c) and (d) the scan width is 1.76 μm.

Subsequently to the force mapping, the area has been imaged in contact mode, which affords a higher pixel resolution (here, 256 pixels per 1.76 μm). The resulting topography and lateral force images are given in Fig. 5(c) and (d), respectively. Regions associated with 2LG display an increased adhesion force (bright areas in Fig. 5(a)) and correspondingly a reduced lateral / friction force (dark area in Fig. 5(d)). Compared to 2LG, 1LG domains show lower adhesion and increased friction force, respectively (see also Fig. S5 in the *Supplementary Information*). Notably, from the step edges displayed in the topography image (Fig. 5(c)) it is impossible to differentiate graphene domain edges from SiC terrace edges. It is well-known that step edges can also occur within a uniform graphene domain, due to the growth process consuming carbon atoms of the SiC substrate and thus causing height variations of its surface[42].

In general, the graphene surfaces were largely comprised of 1LG and 2LG domains. Although the KPFM and adhesion images of Fig. 2 show a small 3LG domain, from the Raman maps measured at several locations across the sample surface there was no evidence for a significant fraction of 3LG (less than 10% of the total area) or MLG domains, largely due to a comparatively low lateral resolution of the Raman system of ~400 nm.

A correlation between adhesion and friction forces is described by the Bowden-Tabor adhesion model for interfacial friction. It states that the friction force scales in a linear manner with the true contact area, $A$, and with the shear strength, $\tau$, as a constant factor[48, 51]. While an AFM tip is sliding across a surface, continuous formation and rupture of "bonds" occurs causing a friction force. In addition, a load dependent contribution, $F_p(L)$, may occur where the sliding tip deforms a soft surface layer such as an OTS monolayer. This second contribution is referred to as ploughing force[48], and the total friction force can be written as $F_f = \tau A + F_p$ . Thus, changes in $F_f$ may result from variations in $\tau$, $A$ and $F_p$. If the shear strength, $\tau$, scales with the adhesion force measured (Fig. 5(a)), then $\tau$ should be lower on 1LG. However, this variation can be counterbalanced or even over-compensated by an increase in $A$ or $F_p$.

Notably, the observation of an increased friction force on 1LG (Fig. 5(d)) is consistent with reports of friction force measurements both in ultrahigh vacuum and under ambient conditions. For graphene as well as other 2D materials, including $MoS_2$, $NbSe_2$ and h-BN, Lee et al.[13] found that the friction force increased as the number of layers decreased. The largest force was measured for 1LG domains and attributed to out-of-plane puckering, which causes an increase in the tip-sample contact area, $A$. Such out-of-plane elastic deformation should be more pronounced for 1LG domains due to their low bending stiffness[13].

In addition, if the ploughing term, $F_p$, encompasses the force needed to move the AFM tip through interfacial layers of water molecules, then variations in friction force may also depend on local variations in water density or in the ordering of water molecules nearby the surface. In particular, on hydrophobic areas lower forces may be needed to displace water molecules, since hydrophobicity tends to be accompanied by a low-density depletion layer, which, in turn, allows water slipping[41, 52, 53]. Thus, the observed friction force contrast (Fig. 5(d)) may also result from spatial variations of the water depletion layer.

Irrespective of the causes of the friction force contrast, Fig. 5 exemplifies that LFM can be very useful in locally differentiating graphene domains from each other. Hence, it is highly recommendable to measure both the adhesion and the friction forces, using the same AFM tip, to enable correlative analysis and to develop a better understanding of the surface morphology and related friction force variations.

(d) Estimation of the adhesion force between 1LG/2LG and a single OTS molecule

Since the measured adhesion force scales with the AFM tip size and thus the number of OTS-graphene "bonds" across the tip-sample contact, it is difficult to compare measurements made using different AFM probes. Normalization can be achieved by calculating the adhesion force per OTS molecule if the tip size is known. The adhesion force per OTS molecule in the contact can be estimated by considering the contact area at the pull-off point. The Johnson-Kendall-Roberts (JKR) theory for mechanical contacts between elastic solids accounts for both compressive and adhesive forces[54]. At the instability point where the AFM tip is pulled off from the surface, the contact radius is given by

$$a_{po}^3 = \frac{3}{2}\frac{\pi w R^2}{K}\,, \qquad (1)$$

where $K$ is the reduced modulus of the tip-sample contact, $R$ is the radius of curvature of the tip apex, and $w$ is the work of adhesion per unit area. The relationship between the latter and the force, $F_{po}$, measured at the pull-off point is:

$$F_{po} = -\frac{3}{2}\pi w R\ . \qquad (2)$$

Importantly, $w$ is related to the surface free energies of the sample-liquid and the tip-liquid interfaces, i.e. $\gamma_{sl}$ and $\gamma_{tl}$. Here, the liquid is de-ionized water. With the free energy $\gamma_{ts}$ of the tip-sample interface, $w$ can be written as

$$w = \gamma_{sl} + \gamma_{tl} - \gamma_{ts}\ \ . \qquad (3)$$

The reduced modulus is related to the Young's moduli, $E_t$ and $E_s$, of the tip and sample materials:

$$K = \frac{4}{3}\left[\frac{1-\nu_1^2}{E_1} + \frac{1-\nu_2^2}{E_2}\right]^{-1}\,, \qquad (4)$$

where $\nu_t$ and $\nu_s$ denote the respective Poisson ratios.

An implication of equation (3) is that for two mating surfaces with identical chemistry, the work of adhesion is $w = \gamma_{sl} + \gamma_{tl} = 2\gamma$, as the interfacial free energy is vanishing. Such a configuration occurred when mapping the OTS functionalized area of the test sample TS1 with an OTS functionalized AFM tip (Fig. 3). As can be seen from the histogram in Fig. 3(d), the mean pull-off force measured on OTS, i.e. the area between the gold squares, was ~3.1 ± 0.6 nN. With $\gamma_{sl}$~ 52 mJ/m$^2$ [23, 50] for the OTS-water interface, an approximate value of ~6.3 ± 1.2 nm is obtained for the tip radius, $R$, which can be written as $R = F_{po}/(3\pi\gamma)$ if $w = 2\gamma$ (see Eq. (2)). This tip radius is in reasonable agreement with a typical value of ~8 nm specified by the manufacturer.

Moreover, the JKR model allows estimation of the average adhesion force between a single OTS molecule on the tip surface and 1LG or 2LG. Using the equations (1), (2) and (4), the contact radius and thus the contact area, $A_{po} = \pi a_{po}^2$, at the pull-off can be calculated if the elastic properties of the tip and sample materials as well as the tip radius, $R$, are known. With a cross-sectional area, $A_{OTS}$, per OTS molecule, the number of OTS molecules in the tip-sample contact is $n_{OTS} = A_{po}/A_{OTS}$ and, thus the pull-off force per OTS molecule is given by

$$F_{OTS} = \frac{F_{po}}{n_{OTS}} = \frac{A_{OTS}}{A_{po}} F_{po} \quad . \tag{5}$$

A typical value of $A_{OTS}$ is 0.43 ± 0.07 $nm^2$ [55].

With the equations (1) and (2), the contact area can be expressed in terms of $F_{po}$, i.e. $A_{po} = \pi \left(\frac{R}{K}|F_{po}|\right)^{2/3}$, and the pull-off force per OTS molecules written as

$$F_{OTS} = \frac{A_{OTS}}{\pi}\left(\frac{K^2}{R^2}|F_{po}|\right)^{1/3} \quad . \tag{6}$$

For a tip radius of $R$ ~6.3 nm (see above), the equations (4) and (6) give an $F_{OTS}$ value of ~(300 ± 30) pN for 1LG and of ~(320 ± 20) pN for 2LG. In Fig. 6, the distributions resulting from the fit curves of Fig. 5(b) are shown, after conversion of the $F_{po}$ values into $F_{OTS}$ values. Comparison of the $F_{OTS}$ values to the total adhesion force, $F_{po}$, measured (Fig. 5(b)), suggests that about 6.4 (1LG) to 7.2 (2LG) OTS-graphene "bonds" were ruptured when detaching the AFM tip from the respective graphene domains.

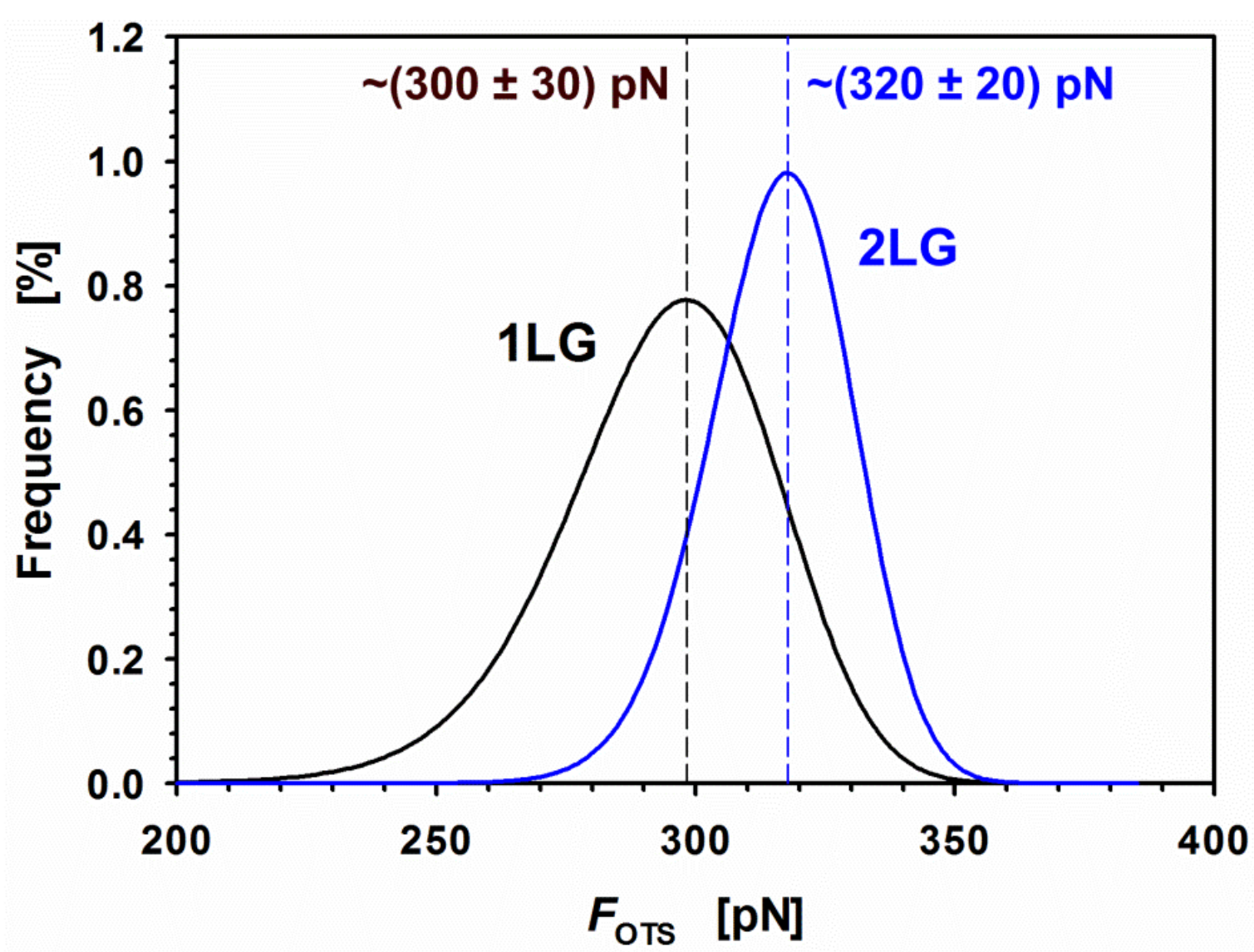

**Figure 6.** Histograms of the pull-off force, $F_{OTS}$, per OTS molecule, calculated by applying the equations (4) and (6) to the deconvoluted peaks of the histogram given in Fig. 5(b). The black and the blue curves show the distributions related to 1LG and 2LG/3LG, respectively.

The given error margins resulted from the width at half maximum of the histograms after conversion of $|F_{po}|$ to $F_{OTS}$ values (Fig. 6). The assumed values of Young's modulus and Poisson's ratio for the OTS functionalized Si AFM tip were 10 GPa and 0.33, respectively[50]. For graphene, the respective values 1000 GPa and 0.15 can be assumed[56, 57]. It should be noted that the calculated values for the adhesion force between a single OTS molecule and the graphene surface are approximate, as the tip radius, $R$, and the area per OTS molecule, $A_{OTS}$, are associated with considerable uncertainties. For $R$~6.3 nm, the uncertainty associated with variations of adhesion force across the OTS-OTS contact (Fig. 3(c) and (d)) is $\Delta R$~ ±1.2 nm. As can be seen from Fig. S6 in the *Supplementary Information*, the resulting uncertainty in $F_{OTS}$ is ~80 pN. Furthermore, the uncertainty in $A_{OTS}$ results from potential imperfections and packing density variations of the OTS monolayer. The $A_{OTS}$ values of ~0.20 [23], 0.21 [25] or 0.43 $nm^2$ [24, 55] (also see references given therein) were derived from monolayers on flat surfaces, whereas the AFM tip presents a highly curved surface which could affect the self-assembly process. Even for the case of a flat substrate, imperfections and associated variations of the packing density of OTS monolayers were reported[47, 48]. Bearing in mind these uncertainties in the absolute values of the adhesion force, $F_{OTS}$, between a single OTS molecule and graphene, it can be concluded that somewhat larger $F_{OTS}$ values have been observed on 2LG domains relative to 1LG domains. Although the values for $F_{OTS}$ are relatively similar for 1LG to 2LG domains (Fig. 6), they are still associated with distinct adhesion force contrasts (Figs. 5(a) and 6(a)) and thus clearly observable under the chosen conditions. Following the rationale of Fig 1(b), which applies to the present case of adhesion measurement in water, a highly polar liquid, this finding indicates that 1LG domains are less hydrophobic than 2LG domains.

## Conclusions

An CFM based approach has been demonstrated that allows differentiation of 1LG and 2LG domains with regard to their degree of hydrophobicity. Using hydrophobic AFM tips functionalized with $CH_3$-terminated silane molecules (OTS), spatial variations of the tip-sample adhesion force have been measured for the case of graphene samples grown epitaxially on SiC. In the presence of water, hydrophobic attraction occurs and a larger adhesion force has been measured on 2LG, as compared to 1LG domains, thus indicating that 2LG domains are more hydrophobic than 1LG domains. The contrast mechanism based on the hydrophobic attraction has also been demonstrated for the case of test samples exhibiting a hydrophobic/ hydrophilic pattern. Thus, this approach involving CFM shows that variations in hydrophobicity occur between 1LG and 2LG. Conversely, under ambient conditions an increased adhesion force was measured on 1LG, due to the prevailing capillary force resulting from the presence of a thin water layer.

Furthermore, by assuming Johnson-Kendall Roberts contact mechanics and relating the measured adhesion force distributions to the tip-sample contact area, approximate values of ~300 and 320 pN for the adhesion force between a single OTS molecule and 1LG and 2LG, respectively, have been calculated. Major sources of uncertainty are the tip size and the average surface area per molecule of the OTS monolayer on top of the highly curved AFM tip. Further to the adhesion force measurements, friction force measurements have shown a good sensitivity for spatial variations in the graphene surface chemistry. Hydrophobic domains with the increased adhesion force were concurrently demonstrating a lower friction force. The results are explained by the fact that different levels of hydrophobicity of 1LG and 2LG are likely to affect the local arrangement of water molecules and, in turn, the sliding motion of the AFM tip.

In general, different levels of hydrophobicity cause variations in the wetting behavior under ambient conditions and are of particular relevance for applications where graphene devices or materials are exposed to non-zero levels of humidity. Measurement of domain-to-domain variations in the degree of hydrophobicity should allow detailed analysis of local doping effects and interfacial chemistry of graphene based composites and nanosensors. Thus, the demonstrated method is a promising approach for the optimization of graphene devices operated under ambient conditions, as their characteristics and performance depend critically on the particulars of the graphene morphology and related variations in the propensity for water adsorption. Future studies could apply the approach demonstrated here to a range of graphene surfaces to elucidate effects of the growth method or device fabrication technique on spatial variations in the degree of hydrophobicity, e.g. to develop a better understanding of wetting behavior variations.

## Materials and Methods

*Graphene synthesis.* The substrates (II-VI, Inc.) were ~8x8 $mm^2$ in size and made of semi-insulating (0001)SiC (resistivity >$10^{10}$ ohm cm), misoriented ~0.05° from the basal plane mainly in the (11-20) direction. Graphene was synthesized via Si sublimation from SiC using an overpressure of an inert gas. The substrates were etched in an $H_2$ atmosphere at 100 mbar using a temperature ramp from room temperature to 1560°C to remove polishing damage. At the end of the ramp, the $H_2$ was evacuated and Ar gas added to a pressure of 100 mbar (the transition takes about 2 minutes). The graphene was then synthesized at 1620°C for 30 min in the Ar. Afterwards, the sample was cooled in Ar to 800°C [4]. Two or more layers of graphene formed in this fashion are known to be Bernal stacked[58].

*Chemical force microscopy (CFM).* Employing a Cypher AFM system (Asylum Research, Santa Barbara, CA), the adhesion forces between chemically functionalized AFM probes and graphene surfaces were measured. The AFM system was fitted with a superluminescent diode to minimize signal oscillations resulting from optical interference of a non-zero fraction of light reflected off the sample surface with the light reflected off the cantilever. The version of the Asylum Research control software was 111111+1719. The instrument was operated in a laboratory controlled to 22 ± 1°C and 44 ± 10% relative humidity. It was calibrated using an NPL traceable blaze grating and a step height standard to ~6% relative uncertainty. The AFM cantilevers used were of type CSC38 with an aluminum reflective coating and a typical tip radius of ~8 nm (Mikromasch Europe, Germany). The optical lever sensitivity and the spring constant for cantilever deflection were measured by running force-displacement curves on a plain Si wafer surface and by recording spectra of the thermally excited cantilever vibrations, respectively. Of the three rectangular silicon cantilevers carried by a chip of type CSC38, cantilevers of ~300 or ~350 μm length were used with measured values of their spring constant in the range of ~80 to 200 pN/nm and ~80 to 150 pN/nm, respectively. To allow hydrophobic behavior of the AFM tip, the cantilevers were functionalized with OTS. Prior to the immersion into a ~$10^{-3}$ mol/L solution of OTS in toluene, the cantilevers were treated in a UV/ozone cleaner for ~20 min. Immediately after the immersion, the cantilevers were washed three times, in chloroform, acetone and again in chloroform. The AFM force measurements were undertaken in de-ionized water, by immersing the cantilever into a drop of ~80 μL volume on top of the graphene sample. The force-displacement curves were run at a scan rate of ~0.81 Hz and a z-velocity of ~1.93 μm/s. They were recorded over arrays of points and the magnitude of the pull-off peaks occurring upon retraction of the probe was analyzed. Typically, the pixel number of the resulting force maps was 64x64. To check the functionalization of the cantilevers and to characterize related adhesion contrasts, test samples with an array of Au squares on top of a $Si/SiO_2$ wafer surface were prepared. The array was fabricated using electron beam lithography (EBL) patterning of a resist layer, sputter coating, and subsequent lift-off. The Au squares measured ~1x1 $\mu m^2$ in size; and the thickness of the Au coating was ~70 nm, on top of a NiCr adhesion layer of ~7 nm thickness. One of the two test samples was functionalized with OTS in the same way as the cantilevers. As the OTS molecules bind to the silica ($SiO_2$) surface of the Si wafer but not to the Au squares, the area between the hydrophilic Au squares is rendered hydrophobic. The second test sample was used with no further functionalization. After removal of carbonaceous contaminations by UV/ozone cleaning, the entire surface was rendered hydrophilic. Indeed, a drop of de-ionized water was found to wet the cleaned surface readily.

*Kelvin probe force microscopy (KPFM) under ambient conditions.* Employing an Icon AFM system (Bruker Nano Surfaces, CA), pull-off forces and the electrical surface potential were measured in PeakForce mode, using highly doped Si probes (Bruker PFQNE-AL) with a spring constant of ∼0.9 N/m. Measurements were carried out using the frequency-modulated KPFM (FM-KPFM) technique. The relative humidity in ambient was ~40%.

*Raman spectroscopy*. Raman intensity maps of the sample were obtained using a 2 mW 532 nm probe confocally focused by a 100X objective resulting in a lateral spatial resolution of (0.4 ± 0.1) μm and the scattered light was analyzed using a Horiba Jobin-Yvon HR800; the spectral resolution was (3.1 ± 0.4) $cm^{-1}$.

*Analysis of AFM data.* The initial processing of force map data was done using the microscope software, version 13.04.77 under IgorPro 6.34A. This part included setting of the calibration parameters and zeroing baseline offsets of force-indentation curves. For analysis of histograms the Scanning Probe Imaging Processor SPIP version 6.1.0 (Image Metrology, Denmark) was used. This step encompassed flattening and definition of masks to select certain areas, such as the gold squares of the test samples. Fitting of individual peaks of a distribution and calculation of the combined distribution were done using SigmaPlot version 12.0 (Systat Software, CA). The LFM images analyzed were calculated from the lateral force images measured in trace and retrace modes by $V_{lat} = (V_{lat}^{trace} - V_{lat}^{retrace})/2$.

**Conflict of Interest**

The authors declare no competing financial interest.

**Acknowledgements**

We are grateful to Tom Wren for microfabrication of the test samples and to Hector Corte Leon for characterization of the probes. This work has been co-funded both by the UK National Measurement System through the Innovation R&D Programme (project Graphene) and by EC grants Graphene Flagship (No. CNECT-ICT-604391). Work at the U.S. Naval Research Laboratory was supported by the Office of Naval Research.

## ***Supplementary Information for:***

# Thickness dependent hydrophobicity of epitaxial graphene

M. Munz[1], C.E. Giusca[1], R.L. Myers-Ward[2], D.K. Gaskill[2], and O. Kazakova[1]

[1]National Physical Laboratory, Hampton Road, Teddington TW11 0LW, UK
[2]U.S. Naval Research Laboratory, Washington, DC 20375 USA

Contents:

(A) Raman confocal spectroscopy maps of epitaxial graphene on SiC

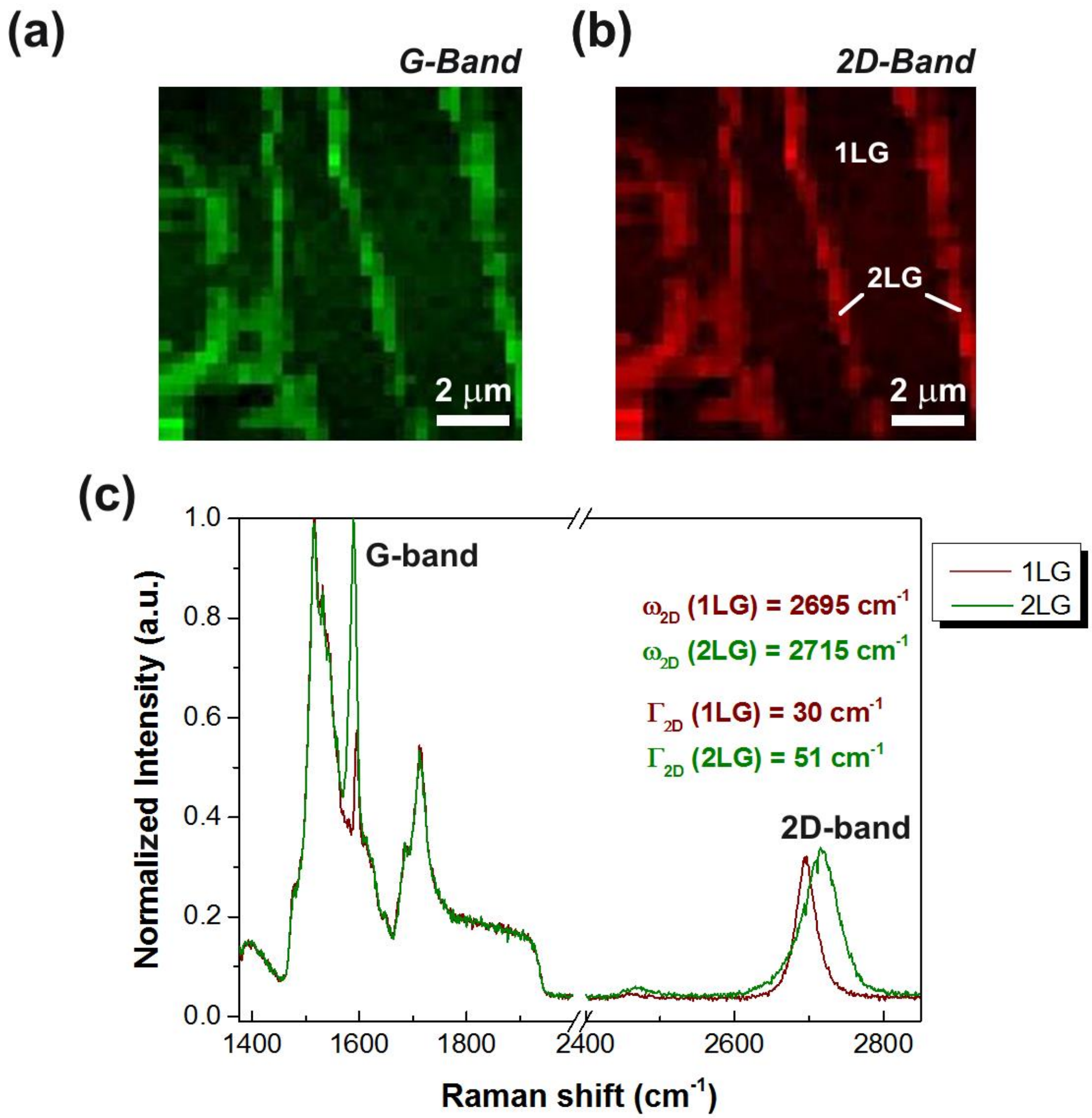

**Figure S1**. (a) G-band and (b) 2D-band Raman intensity maps, obtained using laser light with a wavelength of 532 nm. (c) Representative individual Raman spectra for 1LG and 2LG. In the G-band range the spectra are dominated by Raman modes associated with the supporting SiC substrate. Individual Raman spectra for 1LG and 2LG in (c) display both the G- and 2D modes typical to graphene, as well as the second order features of SiC in the range 1450 - 1750 $cm^{-1}$. Both the G and 2D-band of 2LG display a blue shift compared to that of 1LG, while becoming more asymmetric and almost doubling in width. Changes in the peak positions and widths are only associated with graphene and not with SiC modes. The 2D-band peak position, ω, and the full-width-at-half-maximum, Γ, for 1LG and 2LG, respectively, are given in (c).

(B) Adhesion map of test sample TS2

In the case of test sample TS2 having been exposed to ambient conditions, a larger adhesion force was measured on the Au squares (Fig. S2(c)), presumably due to some carbonaceous contaminations. The observed adhesion contrast suggests that the contaminations adsorbed preferentially on the Au squares and rendered them less hydrophilic than the surrounding silica surface. The related histogram shows distributions around ~2.4 and 9.2 nN (see Fig. S3).
In contrast, no significant adhesion contrast was found if the same sample TS2 was measured soon after UV/ozone cleaning (Fig. S2(d)). In the relative absence of carbonaceous contaminations, the surface was hydrophilic throughout and the hydrophobic attraction vanishing.

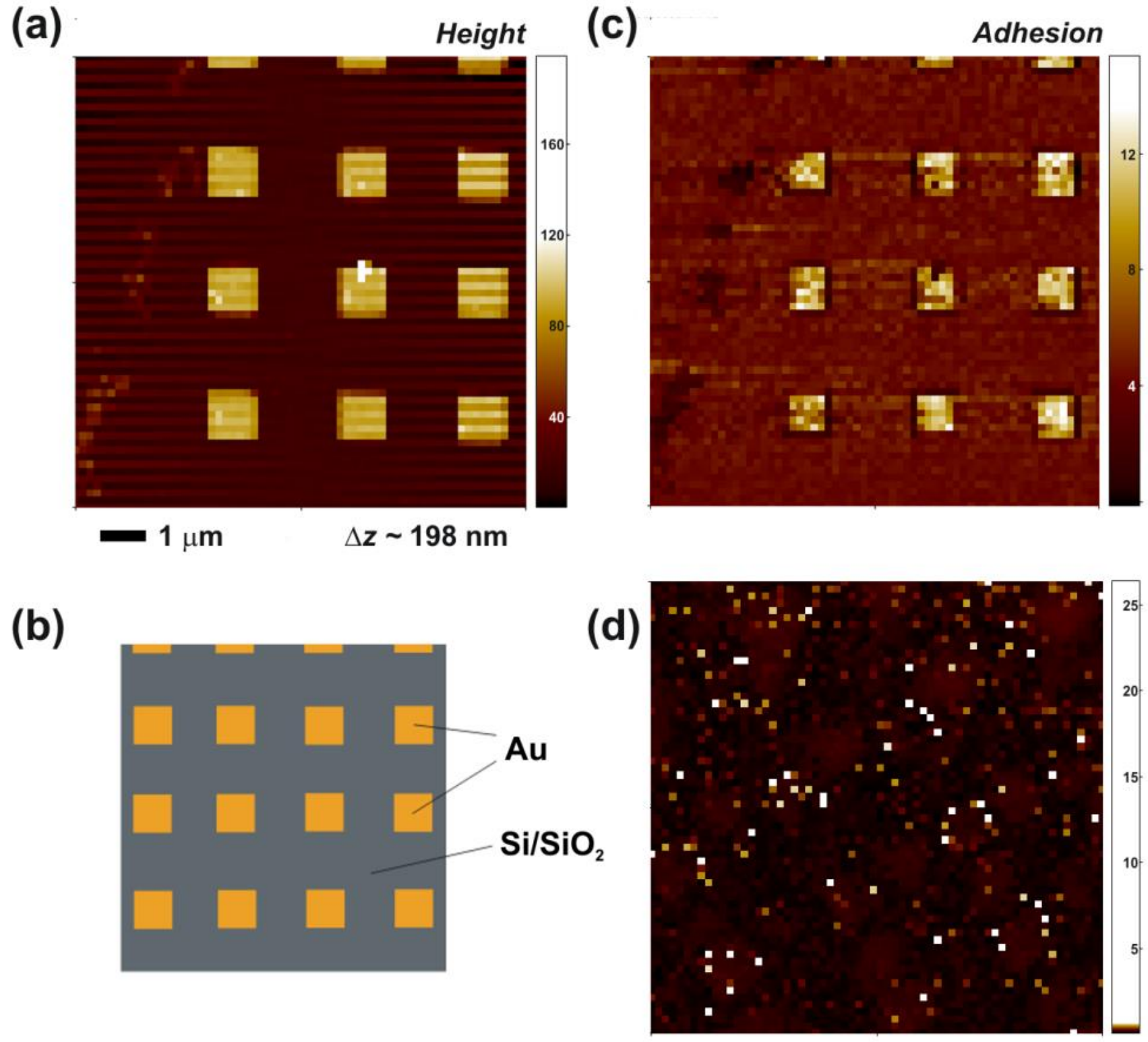

**Figure S2.** Results of force mapping of TS2 (no OTS functionalization), using an OTS functionalized probe. (a) Height map, (b) schematic of the chemical pattern, (c) adhesion map of the sample kept under ambient conditions, (d) adhesion map measured soon after UV/ozone cleaning of the sample. In (d), slight variations in adhesion force level can be seen at positions that are not commensurate with the array of gold squares. These variations are due to weak baseline oscillations caused by some optical interference of laser light reflected off both the cantilever and the sample surface. In (c) and (d) the force range is ~15.5 and 26.3 nN, respectively. In (a, c, d), the scan width is 10 μm, pixel number is 64x64.

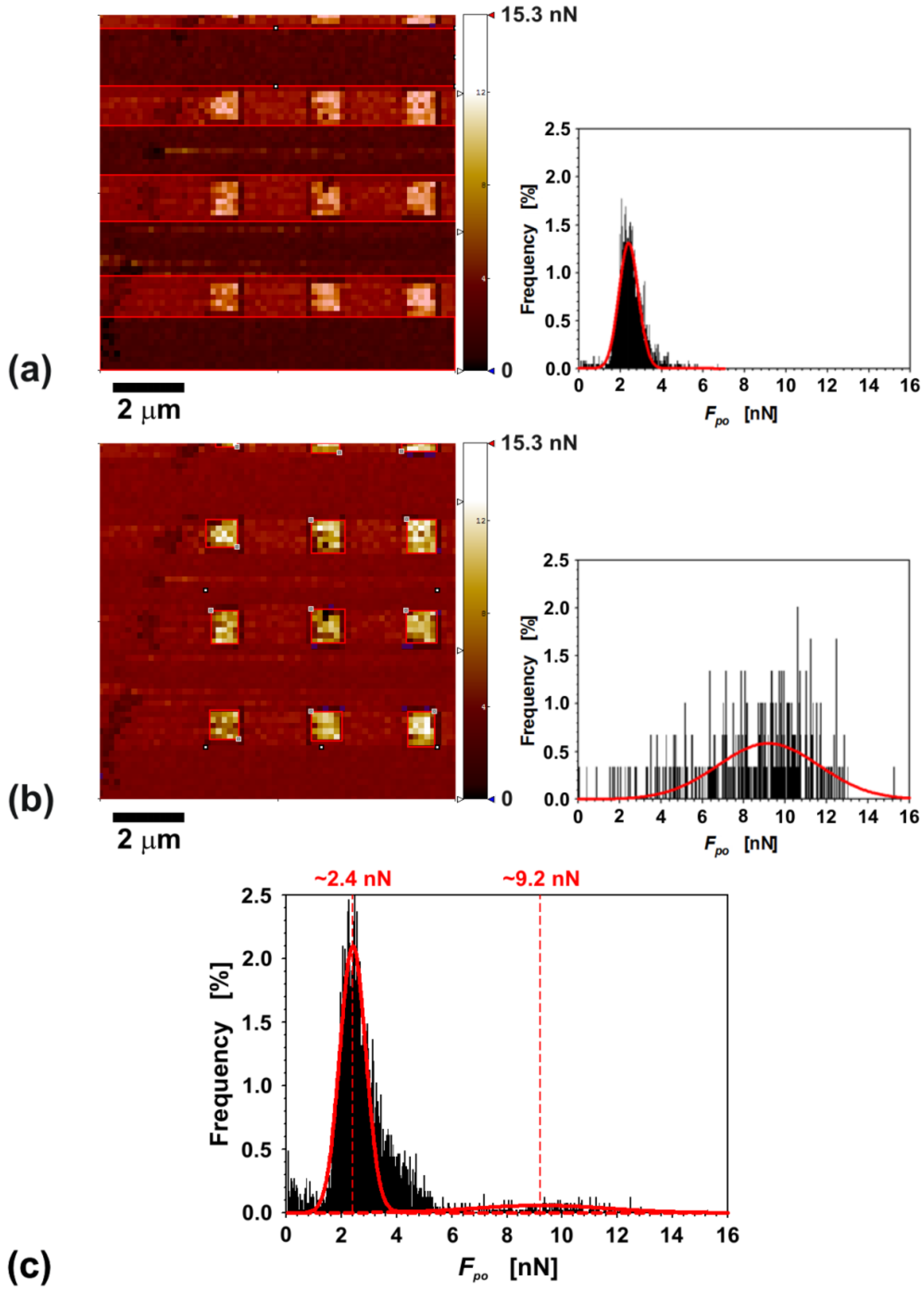

**Figure S3.** Statistical analysis of the adhesion map given in Fig. S2(c). (a) Analysis of the $SiO_2$ surface. It has been selected by a mask and the discarded area is marked with an orange shade. The distribution can be described by a Gaussian, as indicated by a solid red line in the histogram. (b) Analysis of the bright areas selected by a mask. (c) Histogram of the entire map and its approximation by a linear combination of both Gaussians. It should be noted that the shoulder to the right of the peak centered around ~2.4 nN is due to signal variations in the regions between the gold squares. They are slightly brighter than the $SiO_2$ surface area associated with the main peak.

(C) Statistical analysis of the adhesion map given in Figure 5(a)

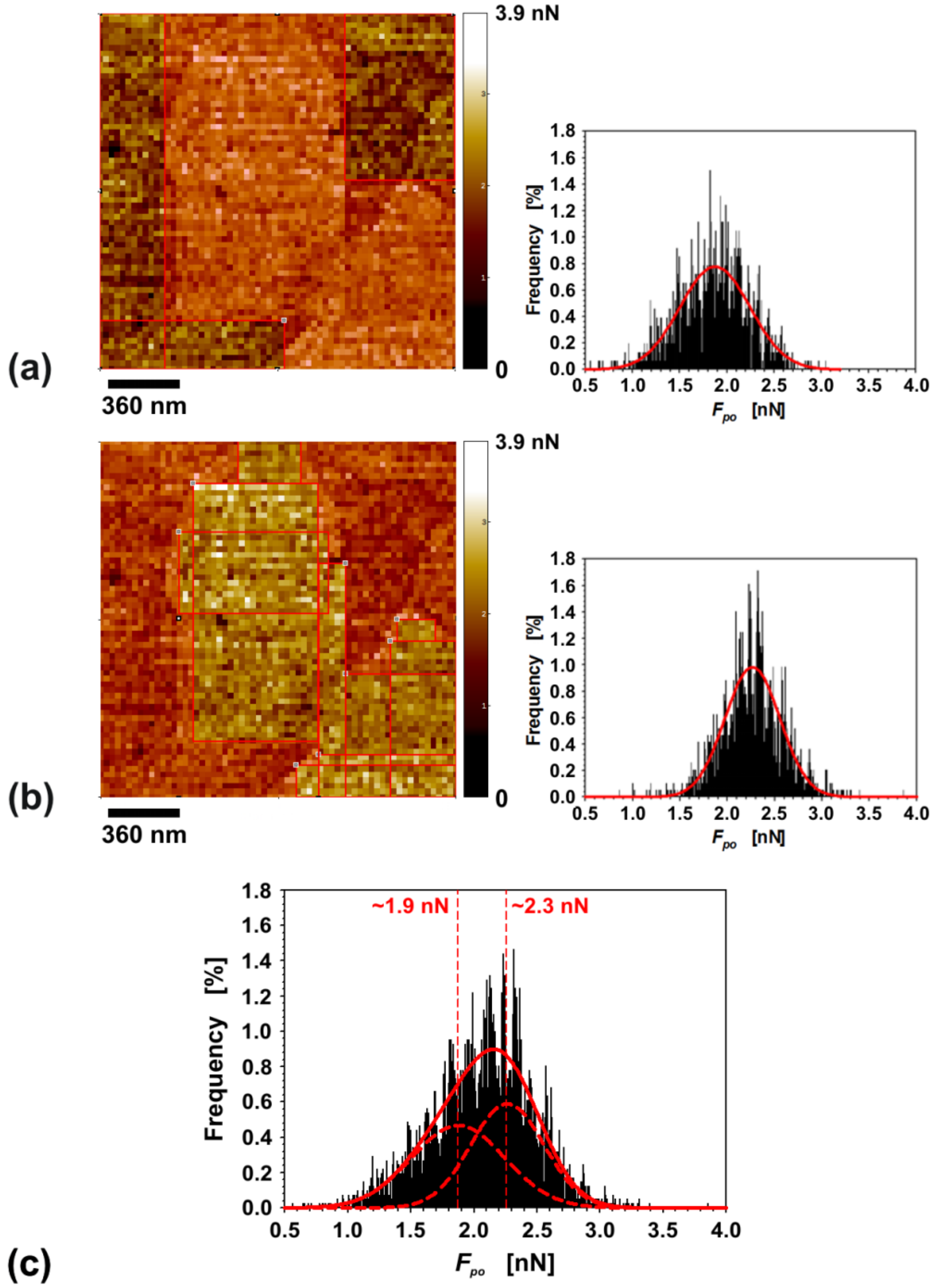

**Figure S4.** Statistical analysis of the adhesion map given in Fig. 5(a). (a) Analysis of the area that shows dark. It has been selected by a mask and the discarded area is marked with an orange shade. The distribution can be described by a Gaussian, as indicated by a solid red line in the histogram. Its center force and FWHM are ~1.9 and ~0.9 nN, respectively. (b) Analysis of the bright areas. The center force and FWHM of the Gaussian describing the distribution are ~2.3 and 0.7 nN, respectively. (c) Histogram of the entire map and its approximation by a linear combination of both Gaussians. The fractional share of the left Gaussian is ~44.2%.

(D) Line 143 of the lateral force image given in Figure 5(d)

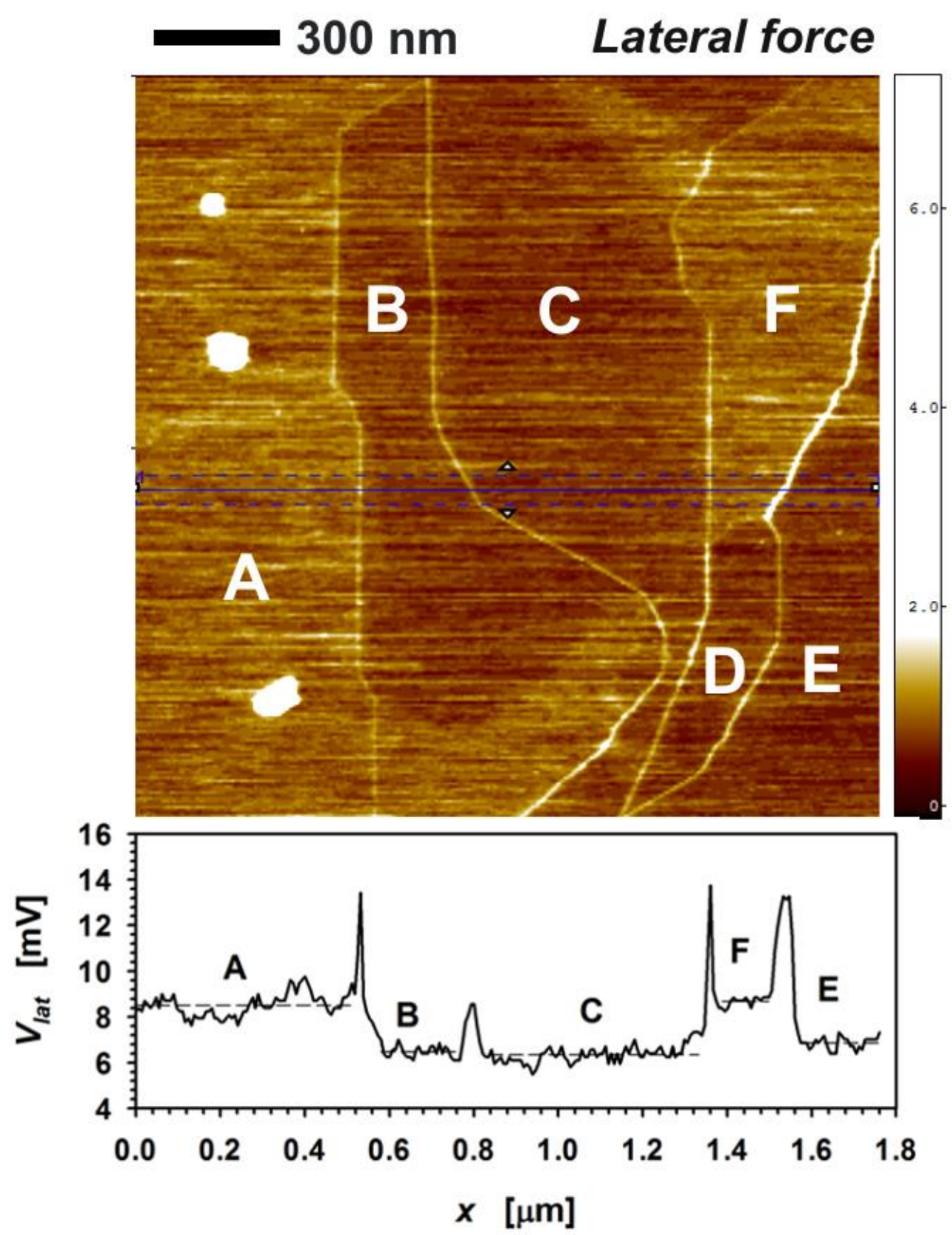

**Figure S5**. Extension of Fig. 5. Lateral force image, as calculated from the lateral force images measured in trace and retrace by $V_{lat} = (V_{lat}^{trace} - V_{lat}^{retrace})/2$, and cross-sectional profile (averaged over 11 consecutive lines within the horizontal rectangle marked with a dashed blue line). Scan width 1.76 µm. For purpose of discussion, the domains are marked with labels B-E.

In the domains B and C the friction force level is about the same as in the domain E. A correlation between the adhesion and the friction forces can also be observed for the domains A and F. Relative to the domains B-E, they show both a lower adhesion force (Fig. 5(a)) and an increased friction force level. Thus, it can be inferred that both domains A and F are comprised of 1LG.

(E) Propagation of uncertainties in the tip radius *R*

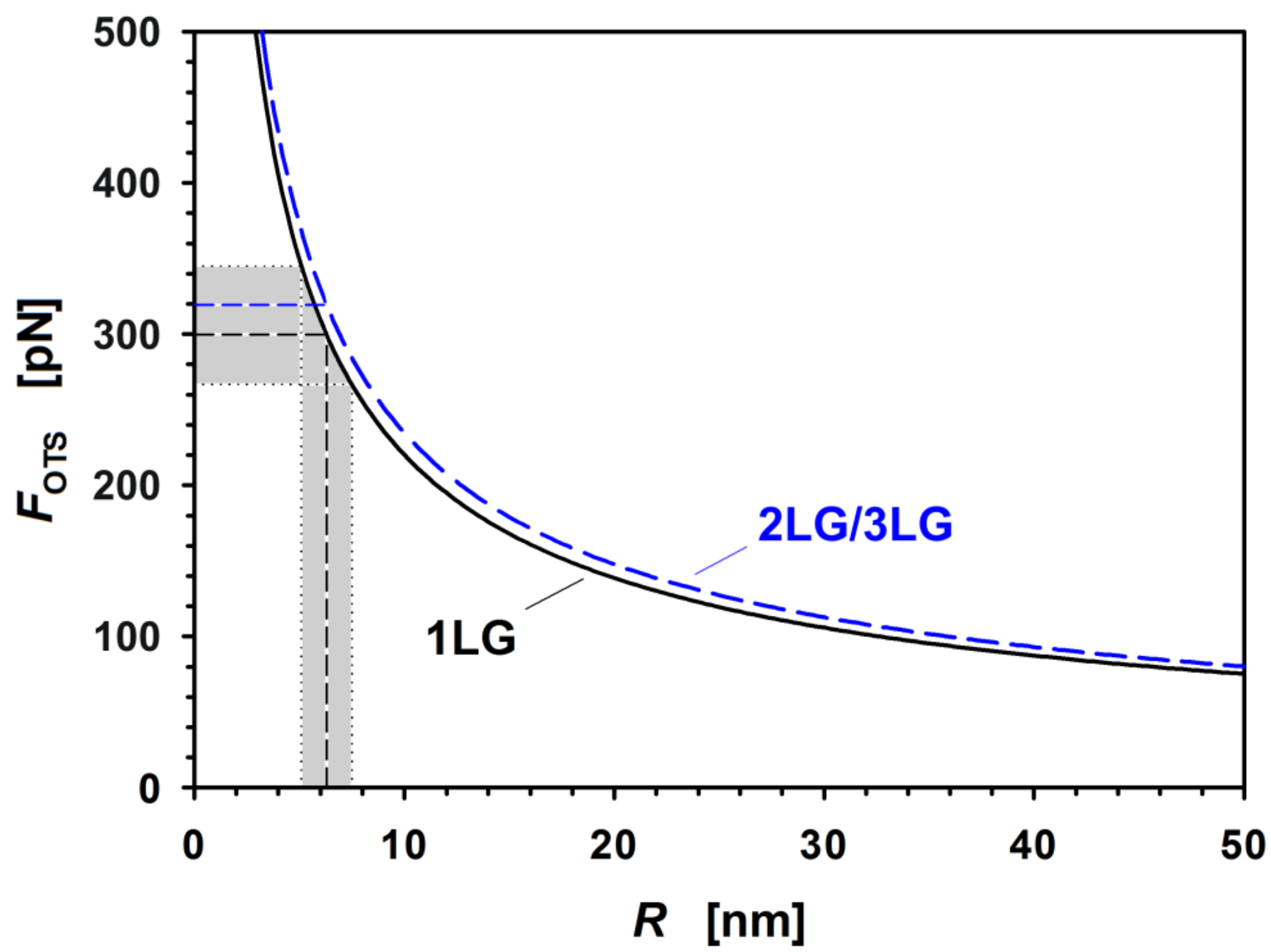

**Figure S6.** Calculated $F_{OTS}$ values for a range of tip radius values and evaluation of the uncertainty in $F_{OTS}$ associated with the uncertainty in $R$. The shaded area illustrates the error propagation for the case of 1LG.

**Abstract graphic**

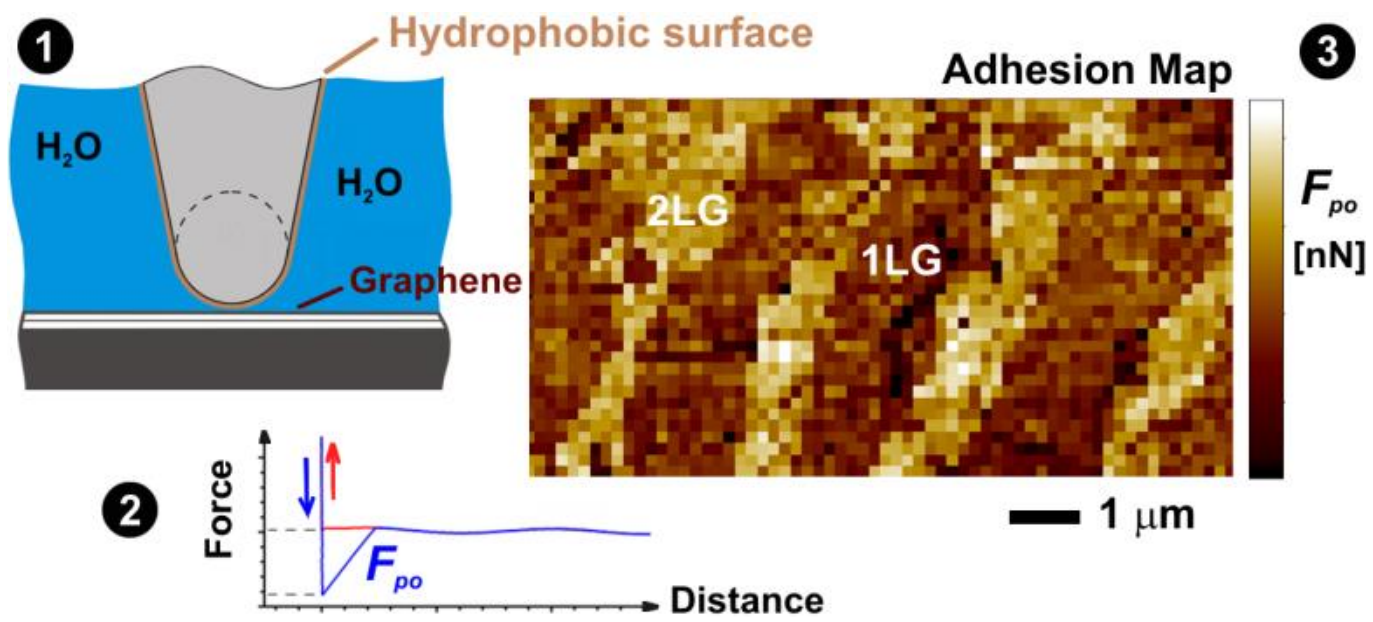